\documentclass[useAMS,usenatbib, 
]{mnras}
\usepackage{natbib}
\usepackage{graphicx}
\usepackage{epsfig}
\usepackage{amssymb}
\usepackage{amsmath}
\usepackage{natbib}
\usepackage{times}

\usepackage[english]{babel}

\title[Interpretation of the settling instability]{Interpretation of the resonant drag instability of dust 
settling in protoplanetary disc}
\author[V.V. Zhuravlev]{V. V. Zhuravlev$^{1}$\thanks{E-mail:
zhuravlev@sai.msu.ru} \\
$^{1}$Sternberg Astronomical Institute, Lomonosov Moscow State University, Universitetskij pr., 13, Moscow 119234, Russia}

\begin{document}

\date{
}

\pagerange{\pageref{firstpage}--\pageref{lastpage}} \pubyear{2017}

\maketitle

\label{firstpage}

\defcitealias{squire-hopkins-2018}{LL05}

\begin{abstract}

The recently discovered resonant drag instability of dust settling in protoplanetary disc is considered as the mode coupling of 
subsonic gas-dust mixture perturbations. This mode coupling is coalescence of two modes with nearly equal phase velocities: 
the first mode is inertial wave having positive energy, while the second mode is a settling dust wave (SDW) 
having negative energy as measured in the frame of gas environment being at rest in vertical hydrostatic equilibrium.
SDW is a trivial mode produced by the bulk settling of dust, which transports perturbations of dust density. 
The phase velocity of SDW is equal to the bulk settling velocity times the cosine of the angle formed by the wave vector
and the rotation axis. 
In this way, the bulk settling of dust makes possible the coupling of SDW with the inertial wave and the onset of the instability. 
In accordance with the concept of the mode coupling, the instability growth rate is proportional to the square root of the 
dispersion equation coupling term, which itself contains the small mass fraction of dust in gas-dust mixture, the squared radial wavenumber of the modes, and the squared bulk settling velocity. 
Thus, the higher is the bulk settling velocity, the heavier clumps of dust can be aggregated by the instability of the same rate.

\end{abstract}

\begin{keywords}
hydrodynamics --- accretion, accretion discs --- instabilities --- protoplanetary discs --- planet formation
\end{keywords}

\section{Introduction}

Relative motion of gas and dust in protoplanetary discs drives the generic dynamical instability of gas-dust mixture 
due to interaction between two components via aerodynamic drag. \citet{squire_2018_0} recognised that instability arises 
each time there is synchronization between the streaming dust motion and the wave of any nature propagating in gas environment.
They termed this type of instability as resonant drag instability. This gave a new look at the previously known 
streaming instability of \citet{youdin-goodman-2005}, which arises due to the radial drift of solids in the midplane of protoplanetary disc.
Currently, the streaming instability is a candidate for the solids growth mediator between pairwise sticking of particles 
and gravitational instability of dust-laden sub-disc via classic Safronov-Goldreich-Ward mechanism, see \citet{safronov-1972} and \citet{goldreich-ward-1973}, which ends with formation of planetesimals.
This view on the role of the streaming instability was established in the work by \citet{johansen-2007-apj-1} and \citet{johansen-2007-apj-2} who investigated its non-linear regime, and subsequently by \citet{johansen-2007-nat} who demonstrated that the resulting dust 
overdensities are gravitationally unstable.
However, the subsequent simulations have shown that this scenario is sensitive to dust fraction: it requires the initial metalicity of disc be a few times higher than the solar metalicity, otherwise turbulence caused by the streaming instability leads to stirring 
rather than clumping of dust, see e.g. \citet{johansen-2009}, \citet{bai-stone-2010}. This constraint becomes more severe due to 
limitations for grain size. 

Along with the known mechanisms of dust pile-up in long-living turbulent structures of disc such as zonal flows and anticyclonic vortices, 
for the review see e.g. \citet{chiang-youdin-2010}, \citet{johansen-2014}, \citet{johansen-2016}, the local amount of dust in a disc 
can potentially be enhanced through a new vigorous instability found by \citet{squire_2018} who termed it as settling instability, 
SI hereafter. Along with the streaming instability, SI is another manifestation of the resonant drag instability caused by sedimentation
of dust. It was revealed that, as compared with the streaming instability, SI growth rate is much higher for small grains with
stopping time shorter than the Keplerian time, because its maximum value does not depend on grain size.
Furthermore, SI operates at larger length-scales. These and other advantages of SI make it a promising candidate for essential link 
in the chain of physical processes leading to planetesimal formation, see the discussion in Section 9.2 of \citet{squire_2018}.


This work is focused on the nature of SI. 
The Lagrangian framework is employed to study the dynamics of small perturbations of gas-dust mixture in two-fluid approximation.
Such a framework is well developed in application to perturbations in single-fluid dynamics. 
In particular, the Lagrangian describing the evolution of perturbations enables to introduce the energy 
of perturbations from the fundamental symmetry of the system with respect to translations in time.
The energy of perturbations is conserved provided that the background flow is stationary, however, it
is not necessarily a positive definite quantity in a moving fluid. The existence of negative energy perturbations
in the flow is well established marker of instability. Indeed, any sink of energy in the system, such as e.g. flux of radiation at 
infinity or viscous damping, provides growth of amplitude of negative energy perturbations. 
Thereby, \citet{friedman-1978b} showed that such perturbations cause the secular instability of rotating stars with respect to gravitational radiation. The energy functional derived in terms of the Lagrangian displacement was referred to as the canonical 
energy, see \citet{friedman-1978}. The concept of the canonical energy has been used in various astrophysical applications
including, for example, stability of relativistic stars, see e.g. \citet{friedman-1998}, stability of magnetic stars, see e.g. 
\citet{andersson-2007}, dynamic tides in stars, see e.g. \citet{ivanov-2010}, \citet{ogilvie-2013}, and many others.


The physical sense of negative energy perturbations is easier to grasp noticing that the total mechanical
energy of the corresponding perturbed flow being itself a positive definite quantity is less than the energy of the background flow.
Probably, for the first time the concept of negative energy in application to space-charge waves in electron beams was 
discussed by L.J. Chu (1951), see the monograph by \citet{briggs-1964} and the book by \citet{pierce-2006}. 
\citet{kadomtsev-1965} were the first to show the existence of negative energy electromagnetic waves in plasma, while the first 
analysis of such possibility for waves in hydrodynamical flows goes back to works by \citet{landahl-1962} and \citet{benjamin-1963}.
A bright exposition on the negative energy waves in context of fluid mechanics and various applications can be found, e.g., in the 
review by \citet{ostrovskiy-1986} and the monograph by \citet{stepanyants-fabrikant-1989}.

One more effect, which provides the growth of amplitude of negative energy wave is coalescence, also termed as coupling, or 
resonant interaction with some other wave having positive energy. 
As soon as the former transfers the energy to the latter, the both waves experience an unbound growth.
By the way, an example of such a coupling of space-charge waves with electromagnetic wave traveling along a helical coil surrounding an electron beam, which is described in the book by \citet{pierce-2006} cited above, 
is quite similar to the subject of this work: roughly speaking, it is enough to replace the electron beam by the stream of the settling dust, the electromagnetic wave in helical coil by inertial wave (IW) in gas environment, the space-charge wave by a settling dust wave (SDW) and finally, the spatial growth by the growth over time. A detailed description of the mode coupling and classification of its variants in application to the streaming of charged particles through a plasma can be found in \citet{briggs-1964}.
As for hydrodynamic applications, a notable work was done by \citet{cairns-1979} who discovered that the mode coupling is responsible for 
a classic Kelvin-Helmholtz instability. Two waves existing on a jump of fluid density in a uniform gravitational field acquire 
energy of different signs and couple with each other, as soon as the layers of different density are in a relative motion.

Another example, where the mode coupling is responsible for hydrodynamical instability, is the well-known Papaloizou-Pringle 
instability of uniform angular momentum rotating tori, see \citet{papaloizou-pringle-1984}, \citet{papaloizou-pringle-1985}, \citet{papaloizou-pringle-1987}, \citet{kojima-1986a}, \citet{kojima-1989} and their citations. 
An exhaustive explanation of Papaloizou-Pringle instability belongs to \citet{glatzel-1987a} and \citet{glatzel-1987b}. 
In a simplified two-dimensional model it was shown that small perturbations grow due to the coupling of surface gravity modes and sound
modes attached to the inner and the outer boundaries of tori.
In the subsequent paper \citet{glatzel-1988} additionally investigated the mode coupling in supersonic plane shear flow.





In this work, the Lagrangian describing the linear dynamics of general two-fluid perturbations is introduced 
for the particular case of dust settling to the midplane of protoplanetary disc. It is found that the energy of certain 
(neutral) mode of such perturbations akin to SDW is negative. SDW itself is a trivial negative energy mode, which represents perturbations 
of dust density transported by the bulk stream of dust through the unperturbed gas environment in the absence of dust back reaction on gas.
Physically, this is the bulk settling of dust that gives rise to a reservoir of negative energy in the form of SDW, from which the 
IW can draw the unlimited amount of energy. 
The latter occurs as soon as the phase velocity of SDW is sufficiently close to the phase velocity 
of IW. As the dust back reaction on gas is taken into account, both modes coalesce with each other giving rise 
to the pair of damping and growing coupled modes with strictly equal phase velocities and the vanished energy, thus, the band of SI appears. The strength of the mode coupling is determined by the coupling term in the dispersion relation, 
which is proportional to the mass fraction of dust in gas-dust mixture.

The study starts from general equations for local dynamics of gas-dust mixture in protoplanetary disc, step by step proceeding 
to a particular model of the perturbed gas-dust mixture with dust bulk settling to the disc midplane. Further, a variational 
principle for evolution of subsonic gas-dust perturbations is formulated and energy of perturbations is derived. 
These results are applied to modes of perturbations in order to demonstrate that SI is caused by the mode coupling between SDW and IW.

\section{General equations for local dynamics of gas-dust mixture in a disc}

In order to consider a small patch inside protoplanetary disc,
the local Cartesian coordinates $x,y,z$ are introduced to represent, respectively, radial $r$, azimuthal $\varphi$ and
vertical $z$ directions. The radial distance is measured ourwards with respect to the host star, while the vertical one is
collinear with disc rotation axis. The centre of a reference frame is located at some point $r_0,\varphi_0, z_0>0$ 
above the disc midplane rotating with angular velocity $\Omega_0$ around the host star.
Accordingly, there are $x\equiv r-r_0,\, y\equiv r_0(\varphi-\varphi_0),\, z\to (z-z_0)$ and it is assumed below that 
$\{x,y,z\}\ll h\ll r_0$, where $h$ is the disc scaleheight. This corresponds to the small shearing box approximation, 
see \citet{goldreich-lynden-bell-1965} and the Appendix A of \citet{umurhan-regev-2004} for detailed derivation of fluid equations.
In this work the same approximation is used to address the two-fluid dynamics of gas-dust mixture.
Both gas and dust are considered as fluids with velocities, respectively, ${\bf U}_g$ and ${\bf U}_p$
measured with respect to some reference velocity ${\bf U}_b$. The latter describes the stationary circular
motion of a single fluid in the gravitational field of the host star,
\begin{equation}
\label{U_b}
{\bf U}_b = -q\Omega_0 x\, {\bf e}_y,
\end{equation}
obeying the following radial and vertical balance 
\begin{equation}
\label{rad_b}
\frac{1}{\rho_b}\frac{\partial p_b}{\partial x} = -\frac{\partial \Phi}{\partial x} + \Omega_0^2 (r_0+x) + 2\Omega_0 U_b,
\end{equation}
\begin{equation}
\label{vert_b}
\frac{1}{\rho_b}\frac{\partial p_b}{\partial z} = -\frac{\partial \Phi}{\partial z},
\end{equation}
where $\Phi$ is the Newtonian point-mass gravitational potential, $p_b$ and $\rho_b$ are pressure and density of the fluid, which can be obtained given the equation of state.
In eq. (\ref{U_b}) the shear rate $q$ is assumed to be constant normally close to the Keplerian value $q=3/2$ in protoplanetary discs.
It can be shown, see the Appendix A of \citet{umurhan-regev-2004}, that in the small shearing box approximation the Euler equation for gas reads 
\begin{equation}
\label{eq_U_g}
\begin{aligned}
(\partial_t - q\Omega_0 x \partial_y) {\bf U}_g - 2\Omega_0 U_{g,y} {\bf e}_x + (2-q)\Omega_0 U_{g,x} {\bf e}_y + \\
({\bf U}_g \nabla){\bf U}_g = - \frac{\nabla p}{\rho_b} + \frac{\sigma}{\rho_b} \frac{\bf V}{\tau},
\end{aligned}
\end{equation}
where $U_{g,x}$ and $U_{g,y}$ are, respectively, $x$- and $y$-projections of ${\bf U}_g$, ${\bf e}_x$ and ${\bf e}_y$ are, 
respectively, $x$- and $y$-orts, while $p$ introduces addition 
to pressure, $p_b$, caused by the influence of dust. The last term in eq. (\ref{eq_U_g}) originates 
from aerodynamic drag of dust particles moving through the gas with the relative velocity ${\bf V}\equiv {\bf U}_p -{\bf U}_g$ 
and mass density $\sigma$. 
The dust particles stopping time $\tau=const$ is a convenient quantity to parametrise aerodynamic drag, see \citet{whipple-1972}.
On the scale much shorter than $h$ the low-frequency dynamics of gas is vortical, thus, eq. (\ref{eq_U_g}) 
is accompanied by the condition of the divergence-free motion
\begin{equation}
\label{eq_rho}
\nabla \cdot {\bf U}_g = 0.
\end{equation}
Note that in the case of $\sigma=0$ equations (\ref{eq_U_g}-\ref{eq_rho}) follow from equations (A.28a-d) of \citet{umurhan-regev-2004}
for unstratified fluid, i.e. provided that their $\rho^\prime$=0.

Similarly, the local dynamics of the pressureless fluid, which mimics the solids suspended in gas environment, is described by 
the following equation
\begin{equation}
\label{eq_U_p}
\begin{aligned}
(\partial_t - q\Omega_0 x \partial_y) {\bf U}_p - 2\Omega_0 U_{p,y} {\bf e}_x + (2-q)\Omega_0 U_{p,x} {\bf e}_y + \\
({\bf U}_p\nabla){\bf U}_p = \frac{\nabla p_b}{\rho_b} - \frac{{\bf V}}{\tau},
\end{aligned}
\end{equation}
where $U_{p,x}$ and $U_{p,y}$ are, respectively, $x$- and $y$-projections of ${\bf U}_p$.
The continuity equation for dust reads
\begin{equation}
\label{eq_sigma}
(\partial_t - q\Omega_0 x \partial_y) \sigma + \nabla (\sigma{\bf U}_p) = 0.
\end{equation}
Again, eqs. (\ref{eq_U_p}-\ref{eq_sigma}) follow from equations (A.15-A.18) of \citet{umurhan-regev-2004} provided that 
their $P=0$, $\rho\to\sigma$ and aerodynamic friction is added in its right-hand side (RHS).
As it should be according to the Newton's third law, the aerodynamic drag of dust acting on gas of unit volume equals to the aerodynamic friction of gas acting on dust of unit volume taken with the opposite sign.

In the case when dust is tightly coupled to the gas it is convenient to rewrite equations in terms of the center-of-mass velocity, see \citet{youdin-goodman-2005}
\begin{equation}
\label{def_U}
{\bf U} \equiv \frac{\rho_b{\bf U}_g + \sigma {\bf U}_p}{\hat\rho},
\end{equation}
where $\hat \rho\equiv \rho_b+\sigma$ is the total density of gas-dust mixture.

Since
\begin{equation}
\label{U_g_V}
{\bf U}_g = {\bf U} - \frac{\sigma}{\hat\rho} {\bf V},
\end{equation}
\begin{equation}
\label{U_p_V}
{\bf U}_p = {\bf U} + \frac{\rho_b}{\hat\rho} {\bf V},
\end{equation}
one arrives at the new equations
\\
\\
\begin{equation}
\label{eq_U}
\begin{aligned}
(\partial_t - q\Omega_0 x \partial_y) {\bf U} - 2\Omega_0 U_y {\bf e}_x + (2-q) \Omega_0 U_x {\bf e}_y + ({\bf U}\nabla) {\bf U} + \\ 
\frac{\rho_b}{\hat\rho} \left \{  \left ( {\bf V} \nabla \left ( \frac{\sigma}{\hat\rho} \right ) \right ) {\bf V} +
2\frac{\sigma}{\hat\rho} \left ( {\bf V} \nabla \right ) {\bf V} \right \} 
= \frac{\nabla p_b}{\rho_b} - \frac{\nabla(p+p_b)}{\hat \rho},
\end{aligned}
\end{equation}

\begin{equation}
\label{eq_V}
\begin{aligned}
(\partial_t - q\Omega_0 x \partial_y) {\bf V} 
- 2\Omega_0 V_y {\bf e}_x + (2-q) \Omega_0 V_x {\bf e}_y + \\ ({\bf U}\nabla) {\bf V}  + ({\bf V}\nabla) {\bf U} + 
\frac{\rho_b}{\hat\rho} ({\bf V} \nabla) \left ( \frac{\rho_b}{\hat\rho} {\bf V} \right ) - \\ 
\frac{\sigma}{\hat \rho} ({\bf V} \nabla) \left ( \frac{\sigma}{\hat\rho} {\bf V} \right ) 
 =  \frac{\nabla (p+p_b)}{\rho_b} - \frac{\hat \rho}{\rho_b} \frac{{\bf V}}{\tau},
\end{aligned}
\end{equation}

\begin{equation}
\label{eq_rho_g}
\nabla \cdot \left ( {\bf U} - \frac{\sigma}{\hat\rho}{\bf V} \right ) = 0,
\end{equation}

\begin{equation}
\label{eq_rho_tot}
(\partial_t - q\Omega_0 x \partial_y) \sigma + \nabla ( \hat\rho {\bf U} ) = 0.
\end{equation}


\subsection{Terminal velocity approximation}

Let the characteristic time- and length-scales of gas-dust mixture dynamics be $t_{ev}$ and $\lambda$, respectively, while 
the dust mass fraction
\begin{equation}
\label{f}
f\equiv \frac{\sigma}{\rho_b} < 1.
\end{equation}
Hereafter, the condition (\ref{f}) implies that gas-dust mixture is not dominated by dust.
If the absolute value of the specific pressure gradient, which governs the dynamics of gas-dust mixture, is
$$
g \equiv \left| \frac{\nabla (p + p_b)}{\hat \rho} \right |,
$$
then the restrictions 
\begin{equation}
\label{tva_1}
\tau_* \equiv \tau \max\{ t_{ev}^{-1},\Omega_0 \} \ll 1,
\end{equation}

\begin{equation}
\label{tva_2}
\lambda^{-1}_* \equiv \frac{g \tau^2}{\lambda} \ll 1,
\end{equation}
greatly simplify equations for gas-dust dynamics. 
More exactly, the terms in the left-hand side (LHS) of eq. (\ref{eq_V}) become small compared to 
the terms in its RHS, while the terms $\sim O(V^2)$ become small compared to the rest terms in LHS of eq. (\ref{eq_U}).
This can be justified as follows.
%
%
%
%

If the restrictions (\ref{tva_1}-\ref{tva_2}) are valid, the specific pressure gradient entering the RHS of eq. (\ref{eq_V}) is balanced by the leading term of this equation, which introduces aerodynamic drag. Dividing eq. (\ref{eq_U}) and eq. (\ref{eq_V}) by $g$ one finds that each of these equations consists of the dimensionless terms of various order in the small $\tau_*$ and $\lambda^{-1}_*$. 
So, the terms in the RHS of eq. (\ref{eq_U}) and eq. (\ref{eq_V}) as well as the gradient term $\sim U^2$ in the LHS of eq. (\ref{eq_U})
are of the zero order in both $\tau_*$ and $\lambda^{-1}_*$. 
The inertial terms $\sim U$ in the LHS of eq. (\ref{eq_U}) differ from unity by factor $\tau_*/\sqrt{\lambda^{-1}_*}$. 
Next, the inertial terms $\sim V$ in the LHS of eq. (\ref{eq_V}) 
are of the order of $\tau_*$, whereas the gradient terms $\sim UV$ in the LHS of eq. (\ref{eq_V}) and the gradient
terms $\sim V^2$ in the both of these equations are, respectively, of the order of $\sqrt{\lambda^{-1}_*}$ and $\lambda^{-1}_*$. 
Additionally, the following order-of-magnitude relations are valid
$$
V \sim g \tau, \quad U \sim \frac{g \tau} {\sqrt{\lambda^{-1}_*}},
$$
so that
$$
V/U \sim \sqrt{\lambda^{-1}_*}.
$$


Hereafter, all terms of the orders of $\tau_*$, $\sqrt{\lambda^{-1}_*}$ and $\lambda^{-1}_*$  
in eqs. (\ref{eq_U}) and (\ref{eq_V}) are omitted, 
which corresponds to what is referred as terminal velocity approximation, 
see \citet{youdin-goodman-2005}. Physically, this means that the inertial forces acting 
on solids in the frame comoving with gas are small compared to drag force and effective gravity force measured in this frame. 
Note that the latter equals to the pressure gradient. As eqs. (\ref{tva_1}) and (\ref{tva_2}) imply, terminal velocity approximation 
is the case for sufficiently small solids with stopping time much shorter than the dynamical time-scale of the problem,
as well as for length-scales of gas-dust mixture dynamics much longer than the solids stopping length $\sim g\tau^2$.

From now on, eqs. (\ref{eq_U}-\ref{eq_V}) read
\begin{equation}
\label{eq_U_2}
\begin{aligned}
(\partial_t - q\Omega_0 x \partial_y) {\bf U} - 2\Omega_0 U_y {\bf e}_x + (2-q) \Omega_0 U_x {\bf e}_y + ({\bf U}\nabla) {\bf U} = \\ 
\frac{\nabla p_b}{\rho_b} - \frac{\nabla(p+p_b)}{\hat \rho},
\end{aligned}
\end{equation}

\begin{equation}
\label{eq_TVA}
\frac{\nabla (p+p_b)}{\hat\rho} = \frac{{\bf V}}{\tau}.
\end{equation}
This work deals with the analysis of the particular linear solution of the set of eqs. (\ref{eq_U_2}), (\ref{eq_TVA}), (\ref{eq_rho_g}) and (\ref{eq_rho_tot}).

\subsection{Equations for perturbations}

The perturbed state of gas-dust mixture can be described by 
the small Eulerian perturbations of the centre-of-mass velocity, ${\bf u}$, the relative velocity, ${\bf v}$, 
the gas pressure, $p^\prime$, and the density of dust, $\sigma^\prime$.  
Those quantities obey the corresponding linear equations

\begin{equation}
\label{pert_U}
\begin{aligned}
(\partial_t - q\Omega_0 x \partial_y)\, {\bf u} - 2\Omega_0 u_y {\bf e}_x + (2-q)\Omega_0 u_x {\bf e}_y + \\ 
({\bf u}\nabla){\bf U} + ({\bf U}\nabla) {\bf u} =  
-\frac{\nabla p^\prime}{\hat \rho} + \frac{\nabla (p+p_b)}{\hat \rho} \frac{\sigma^\prime}{\hat \rho},
\end{aligned}
\end{equation}

\begin{equation}
\label{pert_V}
\frac{\nabla p^\prime}{\hat \rho} - \frac{\nabla(p+p_b)}{\hat\rho} \frac{\sigma^\prime}{\hat\rho} = \frac{{\bf v}}{\tau},
\end{equation}

\begin{equation}
\label{pert_rho}
\nabla \cdot \left ( {\bf u} - \frac{\sigma}{\hat\rho} {\bf v} - 
\frac{\rho_b}{\hat\rho} \frac{\sigma^\prime}{\hat\rho} {\bf V} \right ) = 0,
\end{equation}

\begin{equation}
\label{pert_rho_tot}
(\partial_t - q\Omega_0 x \partial_y)\, \sigma^\prime + \nabla (\sigma^\prime {\bf U} + \hat\rho\, {\bf u})  = 0.
\end{equation}

The set of eqs. (\ref{pert_U}-\ref{pert_rho_tot}) describes the evolution of gas-dust perturbations for any stationary 
solution of eqs. (\ref{eq_U_2}), (\ref{eq_TVA}), (\ref{eq_rho_g}) and (\ref{eq_rho_tot}) on the length-scales much shorter than the scaleheight of a thin protoplanetary disc.
Eqs. (\ref{pert_U}-\ref{pert_rho_tot}) become much more simple for the particular background model considered in the next Section.

\section{Particular case of the dust vertical settling}

In geometrically thin disc
\begin{equation}
\label{g_z}
-\frac{\partial_z p_b}{\rho_b} \equiv g_z \approx \Omega_0^2 z_0, 
\end{equation}
which may be considered as a constant value inside the small box. Note that in the last equality in eq. (\ref{g_z}) 
the Keplerian value of angular velocity, $\Omega_K$, is replaced by $\Omega_0<\Omega_K$. 
This is justified by the small difference between $\Omega_0$ and $\Omega_K$, which is of the order of $(h/r_0)^2$.
It is assumed that the box is located sufficiently far above the disc midplane for $g_z$ to dominate the radial effective gravity, see 
eq. (\ref{rad_b}). The corresponding restriction reads
\begin{equation}
\label{cond_z_0}
\frac{z_0}{r_0} \gg \frac{\Omega_K-\Omega_0}{\Omega_0}.
\end{equation}


In this case the most simple solution of eqs. (\ref{eq_U_2}), (\ref{eq_TVA}), (\ref{eq_rho_g}) and (\ref{eq_rho_tot}) 
describing the dust settling is
\begin{equation}
\label{bg_U}
{\bf U} = 0,
\end{equation}
\begin{equation}
\label{bg_p}
\frac{\nabla (p+p_b)}{\hat\rho} = -g_z {\bf e}_z,
\end{equation}

\begin{equation}
\label{bg_V}
{\bf V} = -\tau g_z {\bf e}_z.
\end{equation}

Finally, eq. (\ref{eq_rho_g}) along with eq. (\ref{bg_V}) implies that
\begin{equation}
\label{bg_sigma}
\sigma=const
\end{equation}
on the local scale considered here.


Once the background model is specified by eqs. (\ref{bg_U}-\ref{bg_sigma}),
one arrives at the following equations for perturbations
\begin{equation}
\label{u}
\begin{aligned}
(\partial_t - q\Omega_0 x \partial_y)\, {\bf u} - 2\Omega_0 u_y {\bf e}_x + 
(2-q)\Omega_0 u_x {\bf e}_y = \\ -\nabla W -\frac{f}{1+f} g_z \delta {\bf e}_z,
\end{aligned}
\end{equation}
\begin{equation}
\label{delta}
(\partial_t - q\Omega_0 x \partial_y)\, {\delta} = -\tau \nabla^2 W + \frac{1-f}{1+f} \tau g_z \partial_z \delta,
\end{equation}

\begin{equation}
\label{div_u}
\nabla \cdot {\bf u} = \frac{\tau f}{1+f} \nabla^2 W - \tau f \frac{1-f}{(1+f)^2} g_z \partial_z \delta,
\end{equation}
where $W \equiv p^\prime/\hat\rho$ and the relative perturbation of the dust density $\delta \equiv \sigma^\prime/\sigma$.
Note that eqs. (\ref{u}-\ref{div_u}) are valid for $f$ not necessarily of the small value.

\subsection{Equations for axisymmetric perturbations in the leading order in small dust fraction}

Let additionally constrain the dust fraction $f\ll 1$ and consider axisymmetric perturbations only.
Eqs. (\ref{u}-\ref{div_u}) yield
\begin{equation}
\label{u_x}
\partial_t u_x - 2\Omega_0 u_y = - \partial_x W,
\end{equation}
\begin{equation}
\label{u_y}
\partial_t u_y + (2-q)\Omega_0 u_x = 0,
\end{equation}
\begin{equation}
\label{u_z}
\partial_t u_z = - \partial_z W - f g_z \delta,
\end{equation}
\begin{equation}
\label{delta_2}
\partial_t \delta = - \tau (\partial^2_{xx} + \partial^2_{zz}) W + \tau g_z \partial_z \delta,
\end{equation}
\begin{equation}
\label{div_u_2}
\partial_x u_x + \partial_z u_z = 0.
\end{equation}

Hence, in this the most simple case the small gas-dust perturbations of dust settling through the static gas environment 
to the disc midplane are described by eqs. (\ref{u_x}-\ref{u_z}, \ref{div_u_2}) for perturbation 
of the centre-of-mass velocity, ${\bf u}$. These equations are identical to equations for single-fluid dynamics 
of vortical perturbations in a rotating plane shear flow, see e.g. equations (6-9) of \citet{umurhan-regev-2004} with $\partial_y=0$, 
except for the term of vertical balance in the RHS of eq. (\ref{u_z}) proportional to perturbation of the dust density. 
At the same time, perturbation of the dust density is the only quantity that additionally describes the dynamics of dust 
by means of eq. (\ref{delta_2}). The perturbed flow of dust is affected by gas through the term $\propto \nabla^2 W$ in the RHS of eq. (\ref{delta_2}). 
In the limit $f\to 0$ the set of eqs. (\ref{u_x}-\ref{div_u_2}) splits into eqs. (\ref{u_x}-\ref{u_z}, \ref{div_u_2}), which 
determine the dynamics of solely the gas perturbations with ${\bf u}$ becoming identical to the velocity of gas,
and eq. (\ref{delta_2}), which separately determines the dynamics of dust perturbations. In the latter situation, solids 
move passively under the action of external gravity and aerodynamic drag. If additionally $\tau\to 0$, 
both the background relative velocity and the perturbation of relative velocity vanish, which results in the conservation of the
density of dust frozen in the divergence-free motion of gas.

The main part of this paper is focused on eqs. (\ref{u_x}-\ref{div_u_2}).

\section{Variational principle for dynamics of small gas-dust perturbations}

It is appropriate to reformulate eqs. (\ref{u_x}-\ref{div_u_2}) in terms of the variables
\begin{equation}
\label{chi}
\chi \equiv \{\varpi,\,\phi,\,u_z,\,\delta\},
\end{equation}
where
$$
\varpi \equiv -\partial_z u_y,
$$
$$
\phi \equiv \partial_z u_x,
$$
It is assumed that components of $\chi$ have non-zero derivatives over $t,x,z$, which are denoted below as $\partial_k\chi_i$.
The usual Einstein's rule of summation over the repeated upper and lower indices is assumed as well. 

First, taking the curl of eqs. (\ref{u_x}-\ref{u_z}) yields the following
\begin{equation}
\label{sys_1}
\partial^2_{tz} u_x - \partial^2_{tx} u_z = 2\Omega_0 \partial_z u_y + f g_z \partial_x \delta,  
\end{equation}
\begin{equation}
\label{sys_2}
-\partial^2_{tz} u_y = (2-q)\Omega_0 \partial_z u_x,
\end{equation}
\begin{equation}
\label{sys_3}
\partial^2_{tx} u_y =  (2-q)\Omega_0 \partial_z u_z.
\end{equation}

Note that eqs. (\ref{sys_1}-\ref{sys_3}) are identical to equations describing 
IW in rigidly rotating fluid provided that $q=f=0$, see e.g. \citet{landau-lifshitz-1987}, paragraph 14. 
The LHS of eqs. (\ref{sys_1}-\ref{sys_3}) are the time derivatives 
of components of vorticity perturbation\footnote{In this work vorticity is a curl of the center-of-mass velocity of gas-dust mixture rather than a curl of the velocity of gas}.

Next, the divergence of eqs. (\ref{u_x}-\ref{u_z}) yields
\begin{equation}
\label{nabla_W}
(\partial^2_{xx} + \partial^2_{zz}) W =  2\Omega_0 \partial_x u_y - fg_z\partial_z \delta.
\end{equation}

Taking additional derivatives over $z$ from equations (\ref{delta_2}), (\ref{nabla_W}), (\ref{sys_3}) 
and employing eq. (\ref{div_u_2}) one comes to the final set of equations for the new variables

\begin{equation}
\label{Sys_1}
\partial_t \phi = \partial^2_{tx} u_z  - 2\Omega_0 \varpi + f g_z \partial_x \delta,  
\end{equation}
\begin{equation}
\label{Sys_2}
\partial_t \varpi = \frac{\kappa^2}{2\Omega_0} \phi,
\end{equation}
\begin{equation}
\label{Sys_3}
\partial^2_{tx}\varpi = - \frac{\kappa^2}{2\Omega_0} \partial^2_{zz} u_z,
\end{equation}
\begin{equation}
\label{Sys_4}
\partial^2_{tz}\delta = \tau g_z \partial^2_{zz}\delta + 2\Omega_0 \tau \partial_x \varpi,
\end{equation}
where $\kappa^2 \equiv 2(2-q)\Omega_0^2$ is the epicyclic frequency squared 
and the term $\propto f\tau$ has been neglected in eq. (\ref{Sys_4}).

The set of eqs. (\ref{Sys_1}-\ref{Sys_4}) can be derived from the requirement that action
\begin{equation}
\label{action}
S = \int {\cal L}(\chi_i, \partial_k \chi_i) d^3{\bf x}\, dt
\end{equation}
with $d^3{\bf x} \equiv dxdydz$ and the Lagrangian density
\begin{equation}
\label{L_full}
{\cal L} = {\cal L}_0 + f {\cal L}_1
\end{equation}
be stationary with respect to arbitrary variations of $\chi_i$ in the small shearing box. 
The subscript `0' denotes the contribution to the full Lagrangian responsible for the dynamics of gas-dust perturbations 
with no account for the drag force acting on gas, i.e. the dust back reaction on gas. In the absence of dust, ${\cal L}_0$ describes single-fluid vortical perturbations in rotating gas. 
The subscript `1' denotes additional contribution to ${\cal L}$ responsible solely for the dynamics of dust in terms 
of the evolution of its density. The first term in ${\cal L}_1$ is the cross-term, which describes both the action of gas on solids and 
the back reaction of solids on gas via the aerodynamic drag. 
If so, eqs. (\ref{Sys_1}-\ref{Sys_4}) are identical to the Euler-Lagrange equations
$$
\frac{\delta {\cal L}}{\delta \chi_i} - \partial_k \left ( \frac{\delta {\cal L}}{\delta (\partial_k \chi_i)} \right ) = 0, 
$$
where explicitly
\begin{equation}
\label{L_0}
{\cal L}_0 = - \varpi \partial_t\phi - \partial_t u_z \partial_x\varpi - \Omega_0\varpi^2 - \frac{\kappa^2}{2\Omega_0}\frac{\phi^2}{2} -
\frac{\kappa^2}{2\Omega_0} \frac{(\partial_z u_z)^2}{2},
\end{equation}
\begin{equation}
\label{L_1}
{\cal L}_1 = g_z \, \left\{ \varpi \partial_x\delta -  
\frac{1}{2\Omega_0} \, \left [ \frac{\partial_t\delta \partial_z\delta}{2\tau} - g_z \frac{(\partial_z\delta)^2}{2} \right ]\, \right\}.
\end{equation}

Note that ${\cal L}$ is known up to an arbitrary (dimensional) constant factor since it describes the linear problem.
The sign of ${\cal L}$ is chosen so that the energy of IW propagating in rigidly rotating fluid 
(i.e. in the case $f=q=0$) be positive, see the next Section.

\subsection{Energy of gas-dust perturbations}

A conserved quantity associated with the invariance of ${\cal L}$ with respect to translations in time is
the energy of perturbations $E \equiv \int {\cal E} d^3{\bf x}$, where
$$
{\cal E} = -{\cal L} + \frac{\delta {\cal L}}{\delta (\partial_t \chi_i)} \partial_t \chi_i
$$
is the energy density. The latter incorporates the basic contribution, 
which is similar to the energy density of gas perturbations in rotating flow in the absence of dust, 
and additional contribution related to perturbations of dust density
\begin{equation}
\label{E_full}
{\cal E} = {\cal E}_0 + f {\cal E}_1.
\end{equation}
Explicitly,
\begin{equation}
\label{E_0}
{\cal E}_0 = \Omega_0\varpi^2 + \frac{\kappa^2}{2\Omega_0} \frac{\phi^2}{2} + \frac{\kappa^2}{2\Omega_0} \frac{(\partial_z u_z)^2}{2},
\end{equation}
\begin{equation}
\label{E_1}
{\cal E}_1 = - g_z\, \left \{ \varpi \partial_x\delta + \frac{g_z}{2\Omega_0} \frac{(\partial_z\delta)^2}{2} \right \}.
\end{equation}

It is straightforward to check that eqs. (\ref{Sys_1}-\ref{Sys_4}) guarantee that $\partial_t {\cal E}$ can be represented as 
the divergence of vector, which is identified with the energy flux vanishing along with perturbations $\chi_i$
as one goes to infinity.
Note that ${\cal E}_0$ is positive definite quantity\footnote{This is true in centrifugally stable flows only}, while in general ${\cal E}_1$ can have any sign depending on the contribution
of the cross term. Nevertheless, the net contribution of the cross term to $E$ can be small or even vanish for certain non-trivial perturbations. Then, owing to perturbations of dust density, ${\cal E}_1$ gives negative amount to the energy.

\section{Modes of gas-dust perturbations}

Let the gas-dust perturbation have the form of plane wave
\begin{equation}
\label{fourier}
\chi_i = \hat \chi_i \exp(-{\rm i}\omega t + {\rm i} {\bf k x}).
\end{equation}
That is, $\hat \chi_i$ are complex Fourier amplitudes of $\chi_i$.
In eq. (\ref{fourier}) ${\bf kx } = k_x x + k_z z$, where $k_x$ and $k_z$ are wavenumbers, respectively, along radial and vertical directions in a disc. The frequency $\omega$ is in general a complex value, so that $\Im[\omega]>0$ implies the exponential growth 
of wave amplitude, i.e. the onset of instability of dust settling to the disc midplane (SI).

Eqs. (\ref{Sys_1}-\ref{Sys_4}) yield
\begin{equation}
\label{mode_eq_1}
-{\rm i} \omega \hat \phi - \omega k_x \hat u_z + 2\Omega_0 \hat \varpi - f g_z {\rm i} k_x \hat \delta = 0,
\end{equation}
\begin{equation}
\label{mode_eq_2}
-{\rm i} \omega \hat \varpi - \frac{\kappa^2}{2\Omega_0} \hat \phi = 0,
\end{equation}
\begin{equation}
\label{mode_eq_3}
\omega k_x \hat \varpi - \frac{\kappa^2}{2\Omega_0} k_z^2 \hat u_z = 0,
\end{equation}
\begin{equation}
\label{mode_eq_4}
\omega k_z \hat \delta + \tau g_z k_z^2 \hat \delta - 2\Omega_0 \tau {\rm i} k_x \hat \varpi = 0.
\end{equation}

\subsection{Energy of mode}

In order to obtain expression for the energy density of mode, $\Re[\chi_i]$ is substituted into eq. (\ref{E_full}) 
and subsequently averaged over the wavelength of mode.
Eqs. (\ref{mode_eq_1}-\ref{mode_eq_4}) provide the following result
\begin{equation}
\label{E_mode}
\begin{aligned}
& \hat {\cal E} = |\hat \varpi|^2 {\rm e}^{2\Im [\omega] t} \left \{ 1 + \frac{(\Re[\omega])^2 + (\Im[\omega])^2}{\kappa^2} \frac{k^2}{k_z^2} + \right . \\ 
&\left . 
\frac{f g_z k_x^2}{(\Re[\omega] + \tau g_z k_z)^2 + (\Im[\omega])^2} \left[ \frac{2\tau}{k_z} (\Re[\omega] +\tau g_z k_z) - 
g_z \tau^2 \right]\,  \right \}, 
\end{aligned}
\end{equation}
where $k^2 \equiv k_x^2 + k_z^2$.

In this way, one obtains the averaged energy density of the mode of gas-dust mixture perturbations for 
a particular wavevector ${\bf k}$, provided that its frequency is known, i.e. there is a solution of a dispersion equation.

\subsection{Dispersion equation}
\label{sec_disp_eq}

As it follows from eqs. (\ref{mode_eq_1}-\ref{mode_eq_4}), $\omega$ obeys equation 
\begin{equation}
\label{disp}
D_g(\omega,{\bf k}) \cdot D_p(\omega,{\bf k}) = \epsilon({\bf k}),
\end{equation}
where 
\begin{equation}
\label{D_g}
D_g(\omega,{\bf k}) \equiv \omega^2 - \omega_i^2,
\end{equation}
\begin{equation}
\label{D_p}
D_p(\omega,{\bf k}) \equiv \omega - \omega_p,
\end{equation}
\begin{equation}
\label{epsilon}
\epsilon({\bf k}) \equiv - f \kappa^2 \omega_p (k_x/k)^2
\end{equation}
with $\omega_i\equiv (k_z/k)\kappa$ and $\omega_p \equiv -\tau g_z k_z$.

\begin{figure}
\begin{center}
\includegraphics[width=8cm,angle=0]{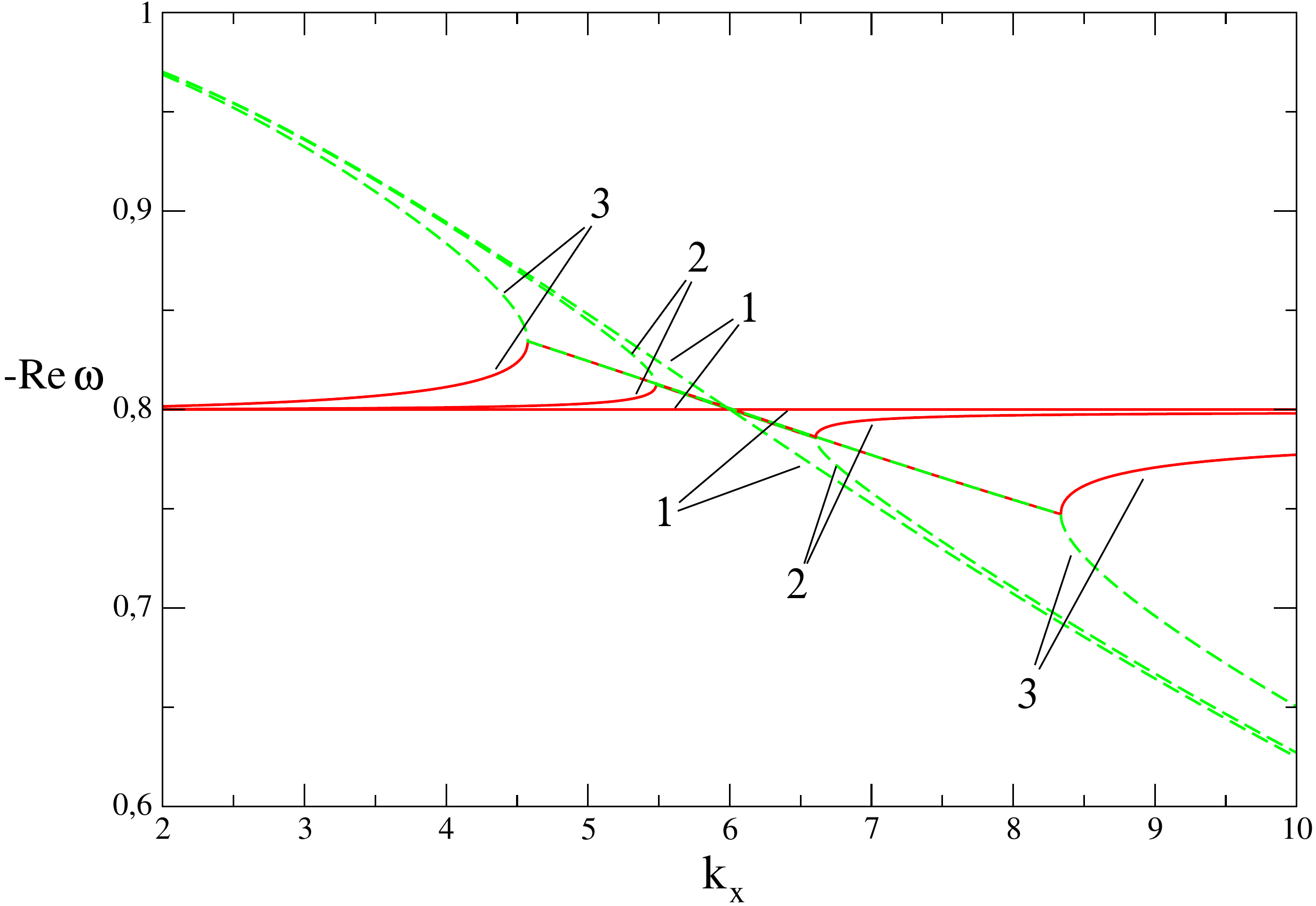}
\end{center}
\caption{The curves show $\Re[\omega]$ taken with the opposite sign, where $\omega$ is the solution of the dispersion equation (\ref{disp}). The parameters are $q=3/2$, $g_z=1 L\Omega_0^2$, $\tau=0.1\Omega_0^{-1}$, $k_z=8 L^{-1}$, where 
$L\lesssim h$ is artificial length-scale. It is assumed that $k_x$ and $\omega$ are measured is units of $L^{-1}$ and $\Omega_0$, 
respectively, while $f=0$ (branches marked as '1'), $f=0.001$ (branches marked as '2') and
$f=0.01$ (branches marked as '3').  In the case $f=0$ the solid and the dashed line correspond to SDW and IW (see text), respectively.
In the case $f>0$ the solid and the dashed lines correspond to modes akin to SDW and IW, respectively.
} \label{fig_1}
\end{figure}

\begin{figure}
\begin{center}
\includegraphics[width=8cm,angle=0]{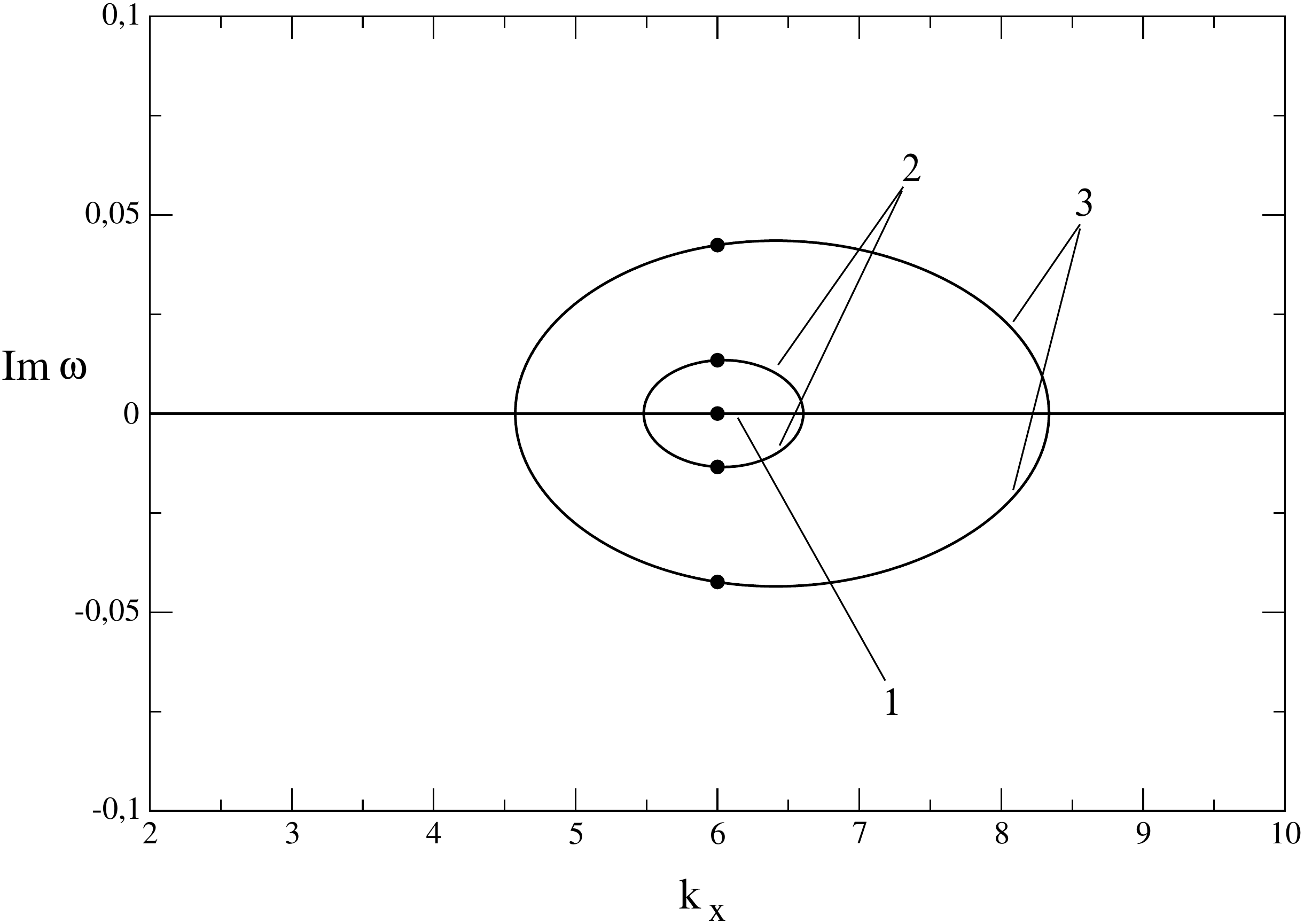}
\end{center}
\caption{Imaginary part of the solution of dispersion equation (\ref{disp}).
Parameters, notations and units are the same as in Fig. \ref{fig_1}.
Filled circles represent analytical solution given by eq. (\ref{approx_im}).
} \label{fig_2}
\end{figure}

\begin{figure}
\begin{center}
\includegraphics[width=8cm,angle=0]{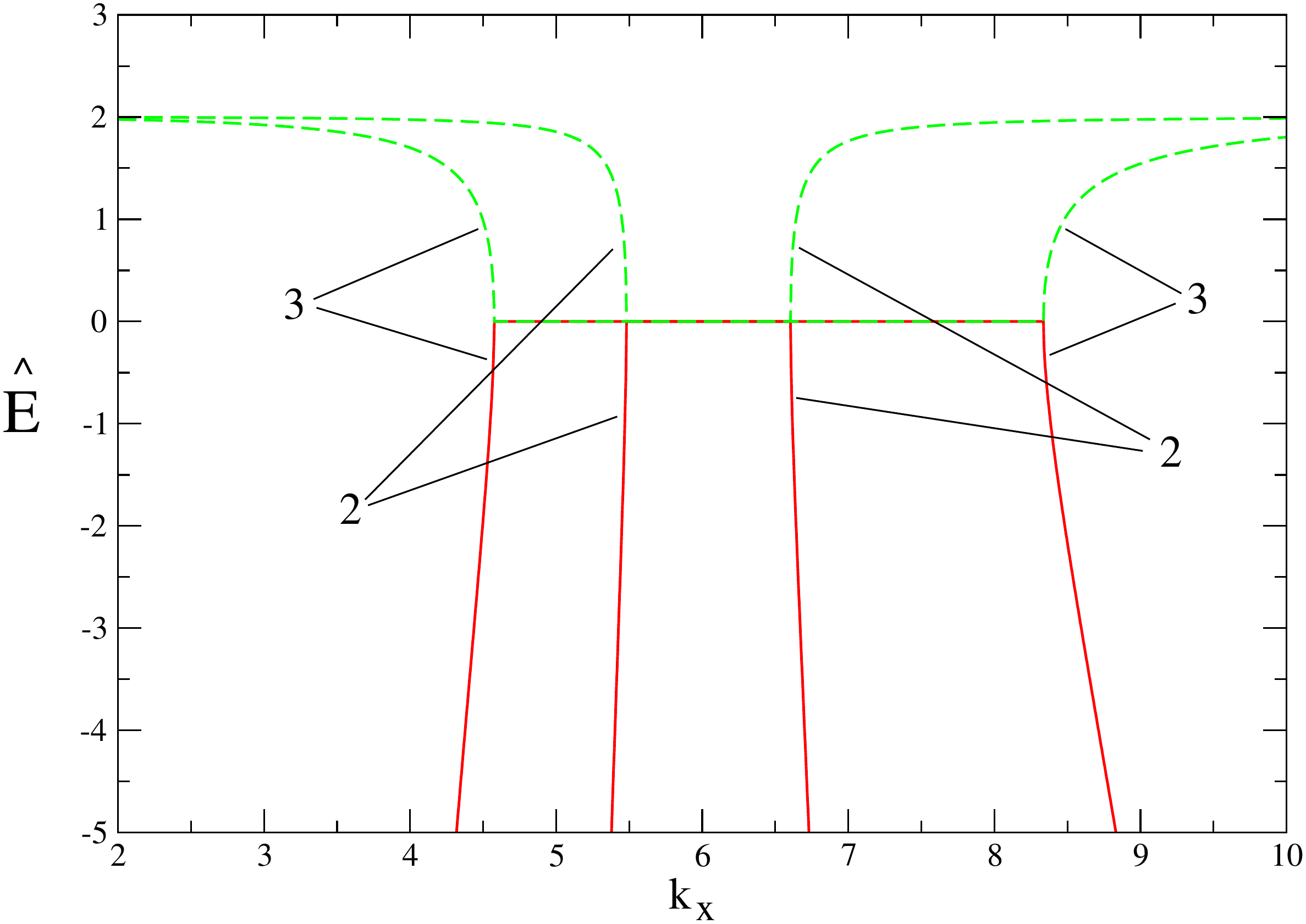}
\end{center}
\caption{Averaged energy density of modes as given by eq. (\ref{E_mode})  for $|\varpi|^2=1$.
Parameters, notations and units are the same as in Fig. \ref{fig_1}, but only the case $f>0$ is shown here. 
The solid and the dashed lines correspond to modes akin to SDW and IW, respectively.}
\label{fig_3}
\end{figure}

An accurate solution of eq. (\ref{disp}) for particular values of parameters is presented in Figs. \ref{fig_1}, \ref{fig_2}.
Note that only two branches of $\omega$ with $\Re [\omega] < 0$ are shown, while there exists a third one with $\Re [\omega] > 0$.
The corresponding value of averaged energy density of mode, $\hat {\cal E}$, normalised by $|\varpi|^2$ is shown in Fig. \ref{fig_3}.

As far as the dust fraction is negligible, $f=0$, there is no instability of gas-dust mixture.
It is notable that for $f>0$ SI sets on in the vicinity of the point, where the frequencies of two branches evaluated
for $f=0$ are equal to each other. For the particular values of parameters taken in Figs. \ref{fig_1}-\ref{fig_3} 
this point corresponds to $k_x=6$. Further, inside the band of SI both modes acquire identical $\Re[\omega]$ (equivalently, phase velocities) and opposite $\Im[\omega]$, i.e. the solution of eq. (\ref{disp}) is given by complex conjugate pair of roots introducing damping and growing modes of gas-dust perturbations.
At the same time, $\hat {\cal E}$ of such perturbations naturally vanishes, since this is a necessary condition for the energy
of each mode to be conserved, see \citet{friedman-1978}.
It is important that outside the band of SI, when $\omega$ is real, the modes have $\hat {\cal E}$ of different signs.

As far as $f=0$, eq. (\ref{disp}) splits into two separate dispersion equations
\begin{equation}
\label{disp_inert}
D_g(\omega_1,{\bf k}) = 0
\end{equation}
and
\begin{equation}
\label{disp_dust}
D_p(\omega_2,{\bf k}) = 0
\end{equation}
describing, respectively, two IW\footnote{Referred to as `epicyclic mode' in \citet{squire_2018}} 
propagating in opposite directions and one else mode, which is referred to as SDW. 
Indeed, eq. (\ref{disp_inert}) can be obtained from eqs. (\ref{mode_eq_1}-\ref{mode_eq_4}) 
setting $\hat \delta=0$. Since the dust fraction is negligible, the remaining variables describe perturbations of gas only. 
The set of equations (\ref{mode_eq_1}-\ref{mode_eq_4}) in this particular limit is identical to those 
that describe IW in rigidly rotating fluid, see e.g. \citet{landau-lifshitz-1987}, paragraph 14.
In astrophysical context, one should take into account the more general case of shearing rotating flow and replace 
$2\Omega$, which is the epicyclic frequency for rigid rotation, by $\kappa$, see e.g. \citet{balbus-2003}, paragraph 3.2.3.
For this reason, below the modes described by eq. (\ref{disp_inert}) are referred as the IW\footnote{Note that according to 
eq. (\ref{pert_V}) such IW produce a non-zero perturbation of relative velocity.}.
On the contrary, eq. (\ref{disp_dust}) follows from eqs. (\ref{mode_eq_1}-\ref{mode_eq_4}) provided that 
$\hat \varpi=\hat \phi = \hat u_z = 0$, which means that gas environment remains unperturbed, whereas arbitrary initial perturbations of dust are transported by the bulk setting of dust with the velocity (\ref{bg_V}). 
The phase velocity of SDW is equal to settling velocity times the cosine of angle formed by the wave vector and the rotation axis.

It is important that one of the branches satisfying eq. (\ref{disp_inert}) crosses the branch satisfying eq. (\ref{disp_dust}) 
on the plane $(\omega,k)$ (or, alternatively, $(\omega,k_{x,z})$) as soon as the condition
\begin{equation}
\label{crossing}
\tau g_z = \frac{\kappa}{k}
\end{equation}
is true. Once (\ref{crossing}) is satisfied, the phase velocities of IW and SDW are equal to each other, 
$\omega_1=\omega_2 \equiv \omega_c = \omega_p$.
Below eq. (\ref{crossing}) is referred to as the condition of a mode crossing. Clearly, the IW participating in the mode crossing propagates downwards to the disc midplane. The second IW propagates upwards to the disc midplane and does not satisfy the condition (\ref{crossing}) for 
any wavevector.

As one can see in Figs. \ref{fig_1},\ref{fig_2}, the non-negligible dust fraction causes the small deviation of 
the branches corresponding to IW and SDW in such a way that in some band around the mode crossing the waves acquire identical 
phase velocities and increment/decrement, which takes maximum value approximately at the mode crossing, compare 
the curves marked by (1) with the curves marked by (2) and (3) in Figs. \ref{fig_1}, \ref{fig_2}. 
Strictly speaking, the waves represented by curves (2) and (3) are the specific kind of waves propagating 
through the gas-dust mixture with dust settling to the disc midplane. Outside the band of instability these waves are akin to IW and SDW 
having, respectively, positive and negative energy densities.
As one approaches the band of instability, the absolute value of the energy density of the both branches vanishes, see Fig. \ref{fig_3}. At the same time, the neutral waves coalesce as seen in Fig. \ref{fig_1} and give birth to a pair of 
growing and damping waves as seen in Fig. \ref{fig_2}. Similar picture occurs in some applications from physics 
of hydrodynamical instability of shear flows, where it is regarded as a mode coupling, see Section \ref{sec_mode_coupling}.

It should be noted that as far as the main restriction of solids tightly coupled to gas, see eq. (\ref{tva_1}), is valid, 
the condition (\ref{crossing}) fits well within the second restriction of the model, see eq. (\ref{tva_2}).
Indeed, in the case of the Keplerian shear eq. (\ref{crossing}) implies that
$$
\lambda \sim g_z \tau \Omega_0^{-1},
$$
which corresponds to length-scale larger than the stopping length of solids settling to the disc midplane\footnote{cf. 
similar restriction obtained by \citet{latter-2011} who considered the streaming instability.} 
$\sim g_z \tau^2$ by factor $\sim (\tau \Omega_0)^{-1} \gg 1$.
The  main restriction (\ref{tva_1}) is valid as far as $\Re[\omega] \sim \Omega_0$ and $\Im[\omega] \lesssim \Omega_0$.

\subsubsection{Approximate solution at the mode crossing}

Eq. (\ref{disp}) is cubic with respect to $\omega$. Seeking the Cardano solution of eq. (\ref{disp}) at the 
mode crossing (\ref{crossing}) one finds that the discriminant 
$$
Q= \frac{8}{27} f \kappa^6 \frac{k_x^2 k_z^4}{k^6} \left ( 1 + \frac{27}{32} f \frac{k_x^2}{k_z^2} \right )
$$ 
vanishes for $f=0$, and eq. (\ref{disp}) acquires a double root. Then, any small non-zero $f>0$ makes $Q$ be a positive
value, which implies that a double root of eq. (\ref{disp}) acquires the adjoint couple of imaginary parts, 
i.e. the instability sets on in some band around the mode crossing.

In the case of small non-zero $f>0$ the Taylor expansion of the Cardano solution of eq. (\ref{disp}) by the orders of $f^{1/2}$ 
at the mode crossing yields the following approximate real and imaginary parts of the frequency
\begin{equation}
\label{approx_re}
\frac{\Re[\omega]}{\omega_c} = 1 + \frac{f}{8} \frac{k_x^2}{k_z^2},
\end{equation}

\begin{equation}
\label{approx_im}
\frac{\Im[\omega]}{\omega_c} = \pm \left ( \frac{f}{2} \right )^{1/2} \frac{k_x}{k_z} \left ( 1 - \frac{5f}{64} \frac{k_x^2}{k_z^2} \right ).
\end{equation}

It is straightforward to check that expressions (\ref{approx_re}) and (\ref{approx_im}) make $\hat {\cal E}$ as given by 
eq. (\ref{E_mode}) vanish for any small $f>0$. Also, eq. (\ref{approx_im}) is in accordance with eq. (5.13) of \citet{squire_2018}
in the leading order in $f$ (their $\mu$). For small value of $f$ used in Fig. \ref{fig_2} the analytical increment/decrement (\ref{approx_im}) virtually coincides with an accurate solution of eq. (\ref{disp}).

\subsection{The mode coupling}
\label{sec_mode_coupling}

The picture revealed in Section \ref{sec_disp_eq} has been examined in the dynamics of waves in shear flows, see \citet{stepanyants-fabrikant-1998}.
\citet{cairns-1979} suggested that the dispersion relation in a form of eq. (\ref{disp}) with small RHS can be interpreted
as the coupling of modes having their own dispersion relations $D_g$ and $D_p$. The coupling causes the deviation of the mode 
frequencies by a small value $\Delta \equiv \omega-\omega_c$ at the mode crossing.
Then,
$$
D_g (\omega) \approx \partial_\omega D_g |_{\omega_c} \Delta, 
$$
$$
D_p (\omega) \approx \partial_\omega D_p |_{\omega_c} \Delta. 
$$
Substitution of these approximations into eq. (\ref{disp}) yields the following solution
\begin{equation}
\label{Delta}
\Delta = \pm \left ( \frac{\epsilon|_{\omega_c}}{\partial_\omega D_g|_{\omega_c} \cdot \partial_\omega D_p|_{\omega_c}} \right )^{1/2},
\end{equation}
which immediately recovers eq. (\ref{approx_im}) in the leading order in $f^{1/2}$ after one employs the eqs. (\ref{D_g}-\ref{epsilon}).
Note that with the help of the mode crossing condition (\ref{crossing}) the coupling term (\ref{epsilon}) can be reduced to the form
$$
\epsilon|_{\omega_c} = -f \omega_p k_x^2 V^2,
$$
so that the increment/decrement of SI becomes
\begin{equation}
\label{Delta_2}
\Delta = \pm {\rm i} \, \sqrt{\frac{f}{2}} |k_x| V.
\end{equation}
Eq. (\ref{Delta_2}) can also be usefull in various estimations related to the problem. By virtue of eq. (\ref{crossing}), 
the maximum growth rate of SI 
$$
\sim \sqrt{\frac{f}{2}}\, \kappa,
$$
i.e. does not depend on the grain size. However, the larger the grains, the
heavier the dust clumps, which can be aggregated by SI.

A more subtle thing is that the sign of the energy of wave is
\begin{equation}
\label{E_sgn}
{\rm sgn} \, \hat E  = {\rm sgn} [ \omega \, \partial_\omega D ],
\end{equation}
where $D(\omega)=0$ is a dispersion equation of wave, see \citet{landahl-1962} and also \citet{cairns-1979} for first
discussion of this feature in hydrodynamics. In the next Section it is confirmed in application to the dynamics of gas-dust mixture.
Since in the considered situation $\omega_p<0$, the coupling term is positive at the mode crossing $\epsilon|_{\omega_c}>0$, 
which implies that the instability occurs as soon as denominator in eq. (\ref{Delta}) is negative, 
which in turn is true provided that the coupling waves have the energy of different signs. 
This allows for clear interpretation of SI as the resonant coalescence of waves in the gas-dust mixture, 
which provides an exchange with energy between the waves. The total energy of coupled mode formed by the coalesced waves 
remains constant, while the non-zero energy flux between the coalesced waves provides the growth of their amplitudes 
as soon as the energy flows from the negative energy wave to the positive energy wave. 
On the contrary, if the energy flows in the opposite direction, a coupled mode is damping.
The two cases described above, represent the pair of complex conjugate solutions of dispersion equation (\ref{disp}) discussed
in Section \ref{sec_disp_eq}.

\subsection{Averaged variational principle}

For the class of neutral modes, when $\Im [\omega]=0$ and amplitudes are constant it is possible to formulate 
the more particular variational principle describing the dynamics of small perturbations, see \citet{whitham}.

Let the modal solution have the form
\begin{equation}
\label{real_mode}
\begin{aligned}
\varpi = \tilde \varpi \cos \theta, \\
\phi = \tilde \phi \sin \theta, \\
u_z = \tilde u_z \cos \theta, \\
\delta = \tilde \delta \sin \theta, \\
\end{aligned}
\end{equation}
where tilded quantities are real and the phase $\theta \equiv -\omega t + {\bf k x}$. 
%
The tilted quantities in eq. (\ref{real_mode}) satisfy the set of eqs. (\ref{mode_eq_1}-\ref{mode_eq_4}), where
the following replacement is made
$$
\hat \varpi \to \tilde \varpi,\, \hat \phi \to -{\rm i}\tilde \phi,\, \hat u_z \to \tilde u_z, \, \hat \delta \to -{\rm i} \tilde \delta. 
$$

According to the averaged variational principle, see \citet{whitham}, there is a Lagrangian 
$$
\bar L = \bar L(\tilde\chi_i, \partial_i\theta),
$$
where $\tilde\chi_i \equiv \{ \tilde\varpi,\tilde\phi,\tilde u_z,\tilde \delta \}$ and 
$\partial_i \equiv \{ \partial_t, \partial_x, \partial_z \}$, 
which generates the set of eqs. (\ref{mode_eq_1}-\ref{mode_eq_4}) for modes of perturbations provided that the action
similar to (\ref{action}) be stationary with respect to arbitrary variations of $\tilde\chi_i$ and $\theta$.
It reads
\begin{equation}
\label{aver_L}
\begin{aligned}
\bar L = & \left \{ \omega \tilde \varpi \tilde \phi + \omega k_x \tilde \varpi \tilde u_z - \Omega_0 \tilde \varpi^2 - 
\frac{\kappa^2}{2\Omega_0} \frac{\tilde \phi^2}{2} -  \frac{\kappa^2}{2\Omega_0} \frac{k_z^2 \tilde u_z^2}{2}  \right \} + \\ 
& f g_z \left \{ k_x \tilde\varpi\tilde\delta + \frac{1}{2\Omega_0} \left [ \frac{\omega k_z}{\tau} \frac{\tilde \delta^2}{2} + g_z k_z^2 \frac{\tilde\delta^2}{2} \right] \right \}.
\end{aligned}
\end{equation}
It can be checked that eqs. (\ref{mode_eq_1}-\ref{mode_eq_4}) are equivalent to the Euler-Lagrange equations 
\begin{equation}
\label{aver_L_eq}
\frac{\partial \bar L}{\partial \tilde\chi_i} = 0.
\end{equation}
It is straightforward to check that equations (\ref{aver_L_eq}) yield an alternative form of the dispersion equation 
(cf. eq. (\ref{disp})), which is nothing but
\begin{equation}
\label{disp_L}
\bar L=0.
\end{equation}
At the same time, the remaining Euler-Lagrange equation
$$
\partial_i \frac{\partial \bar L}{\partial (\partial_i\theta)} = 0
$$
is satisfied identically since $\tilde\chi_i$ and $\omega$, $k_x$, $k_z$ are assumed to be constants.

Finally, the energy of neutral mode following from the averaged variational principal is explicitly
\begin{equation}
\label{aver_E}
\begin{aligned}
\bar E = \frac{\partial \bar L}{\partial (\partial_t \theta)} \partial_t \theta = \omega \frac{\partial \bar L}{\partial \omega} = \\
\frac{\Omega_0 \tilde \varpi^2}{\kappa^2} \frac{k^2}{k_z^2} \omega \frac{\partial D_g}{\partial \omega} + 
f g_z \frac{k_z}{2\tau\Omega_0}\frac{\tilde\delta^2}{2} \omega \frac{\partial D_p}{\partial \omega}.
\end{aligned}
\end{equation}
Note that eq. (\ref{aver_E}) is identical to eq. (\ref{E_mode}) provided that $\Im[\omega]=0$. 
There is no cross term $\propto \tilde \varpi \tilde \delta$ in eq. (\ref{aver_E}) which leads to the conclusion that 
any perturbation of dust density gives negative amount to the energy of neutral mode with $\omega<0$.
In the limit $f\to 0$ 
the first term in eq. (\ref{aver_E}) introduces the energy 
of IW, whereas the last term therein introduces the energy of SDW, which confirms the rule (\ref{E_sgn}) employed
for interpretation of SI, see Section \ref{sec_mode_coupling}.
Clearly, in the case of dust settling to the disc midplane with the velocity (\ref{bg_V})
the mode crossing occurs at negative frequency $\omega_c=\omega_p<0$, therefore SDW has negative 
energy, while the energy of IW remains to be positive. 
As the dust fraction becomes non-negligible, whether the energy of the particular mode of gas-dust perturbations is 
either positive or negative
is determined by the balance of terms in eq. (\ref{aver_E}) for the corresponding solution of eq. (\ref{disp}).
Accurate computation demonstrates that the modes akin to IW and SDW have, respectively, positive and negative energies, see
the branches (2) and (3) in Fig. \ref{fig_3} outside the band of SI.

\section{Conclusions}

The Lagrangian perturbation theory for local dynamics of gas-dust mixture with dust settling to protoplanetary disc midplane is constructed in order to show the existence of negative energy wave in such a flow. 
In general, the negative energy wave is akin to SDW, while it becomes identical to SDW in the absence of dust back reaction on gas.
The dispersion equation for gas-dust perturbations with the account for dust back reaction on gas is reduced to the form typical for 
systems with the mode coupling. In the absence of the coupling term the dispersion equation splits into two independent equations describing one SDW and two IW, respectively. The dispersion curve of SDW crosses the dispersion curve of IW propagating downwards to the disc midplane at the point referred to as the mode crossing. It is revealed that SI occurs due to the energy transfer from (negative energy) SDW to (positive energy) IW in the vicinity of the mode crossing, where the phase velocities of both waves are equal to each other. The coupled modes are represented by the complex conjugate pair of frequencies, for which the analytical expressions are 
derived up to the next order in $f$, see eqs. (\ref{approx_re}) and (\ref{approx_im}). It is checked that the energy of 
the coupled modes vanishes. 

The band of instability expands in the space of wavevectors as the coupling term in the dispersion equation, 
which is proportional to the dust mass fraction, becomes larger. 
The coupling term originates from the product of the last term in eq. (\ref{mode_eq_1}) and
the last term in eq. (\ref{mode_eq_4}). In turn, the former comes from the last term in RHS of eq. (\ref{u_z}), which is proportional
to vertical gravitational acceleration in a disc. Physically, this term is the deviation of the weight of solids acting on gas, which
emerges due to perturbation of the dust density. At the same time, the last term in eq. (\ref{mode_eq_4}) comes from the first term in
RHS of eq. (\ref{delta_2}), which exists as soon as the flow is rotational, see eq. (\ref{nabla_W}). 
Indeed, if it were not for the rotation, the pressure maxima would be absent in the perturbed gas environment. In the linear problem, 
the subsonic pressure maxima, which are necessary both for existence of IW and for dust clumping, 
arise due to the action of Coriolis force. The similar situation was noted by \citet{latter-2011} in context of the streaming instability,
see their discussion about the geostrophic balance. The models that introduce SI and the streaming instability differ from each other 
only by the direction of the effective gravity and the relative velocity in the background solution. 
Presumably, the streaming instability has the similar 
physical interpretation as given above for SI, however, the details are to be revealed in future work.

As one follows derivation of eq. (\ref{crossing}) starting from the general equations (\ref{pert_U}-\ref{pert_rho_tot}), it becomes 
clear that the LHS of eq. (\ref{crossing}) is nothing but the bulk settling velocity of dust given by eq. (\ref{bg_V}), 
also cf. eq. (4.4) of \citet{squire_2018}.
Therefore, the latter defines the characteristic wavelength of the mode coupling and the band of SI.
As has been already recognised by \citet{squire_2018}, the growing gas-dust perturbations of SI have larger length-scales
than that of the streaming instability. Clearly, this is because in the geometrically thin disc settling velocity is higher 
than the velocity of radial drift by factor $\sim (h/r_0)^{-1}$.


The relevance of the streaming instability to planetesimal formation has being extensively studied 
via local simulations of the non-linear dynamics of perturbations in dust-laden disc midplane, see the recent results by \citet{johansen-2017}. 
Although the numerical models usually incorporate the dust settling, SI has not been found so far, see the discussion 
by \citet{squire_2018}. Besides, SI must be less sensitive to external turbulence inherent in astrophysical discs, see
the recent numerical study of dust settling through turbulent gas by \citet{lin-2019}. 
On the contrary, enhanced effective viscosity produced by turbulence may cause an additional non-resonant growth of 
negative energy mode akin to SDW outside of the mode coupling.

At last, it is highly likely that the other resonant drag instabilities revealed by \citet{squire_2018} are caused by coupling of SDW 
with waves of the other origin, such as sound waves and internal gravity waves ubiquitous in protoplanetary discs.
These issues are relegated to future work.


\section*{Acknowledgments}

The author acknowledges the support from the Program of development of M.V. Lomonosov Moscow State University (Leading Scientific School 'Physics of stars, relativistic objects and galaxies').

\bibliography{bibliography}



\end{document}